\documentclass[]{aastex631}

\shorttitle{Formation of CME and Post-eruptive Flow}
\shortauthors{Slemzin et al.}
\graphicspath{{./}{figures/}}

\begin{document}

\title{Formation of Coronal Mass Ejection and Post-eruption Flow of Solar Wind on 2010 August 18 event \footnote{Released on ,,}}

\correspondingauthor{Vladimir Slemzin}
\email{slemzinva@lebedev.ru}

\author[0000-0002-5634-3024]{Vladimir Slemzin}
\affiliation{P.N.~Lebedev Physical Institute of the Russian Academy of Sciences (LPI),53 Leninskiy Prospekt,119991, Moscow, Russia}
\author[0000-0001-9257-4850]{Farid Goryaev}
\affiliation{P.N.~Lebedev Physical Institute of the Russian Academy of Sciences (LPI),53 Leninskiy Prospekt,119991, Moscow, Russia}
\author[0000-0002-5874-4737]{Denis Rodkin}
\affiliation{P.N.~Lebedev Physical Institute of the Russian Academy of Sciences (LPI),53 Leninskiy Prospekt,119991, Moscow, Russia}

\begin{abstract}

The state of the space environment plays a significant role for forecasting of geomagnetic storms produced by disturbances of the solar wind (SW). Coronal mass ejections (CMEs) passing through  the heliosphere often have a prolonged (up to several days) trail with declining speed, which affects propagation of the subsequent SW streams. We studied the CME and the post-eruption plasma flows behind the CME rear in the event on 2010 August~18 observed in quadrature by several space-based instruments. Observations of the eruption in the corona with EUV telescopes and coronagraphs revealed several discrete outflows followed by a continuous structureless post-eruption stream. The interplanetary coronal mass ejection (ICME), associated with this CME, was registered by PLAsma and SupraThermal Ion Composition (PLASTIC) instrument aboard the Solar TErrestrial RElations Observatory (STEREO-A) between August 20, 16:14~UT and August~21, 13:14~UT, after which the SW disturbance was present over 3 days. Kinematic consideration with the use of the gravitational and Drag-Based models has shown that the discrete plasma flows can be associated with the ICME, whereas the post-eruption outflow was arrived in the declining part of the SW transient. We simulated the Fe-ion charge distributions of the ICME and post-CME parts of SW using the plasma temperature and density in the ejection region derived from the Differential Emission Measure analysis. The results demonstrate that in the studied event the post-ICME trailing region was associated with the post-eruption flow from the corona, rather then with the ambient SW entrained by the CME.

\end{abstract}


\section{Introduction} \label{sec:intro}

Interplanetary plasma consisting mainly of the slow moving solar wind (SW) represents a medium, in which recurrent disturbances of SW associated with fast quasistationary streams from coronal holes and non-periodic disturbances associated with coronal mass ejections (CMEs) propagate \citep{lust1963,russell2001}. Depending on their magnitude and direction, CMEs may give rise to geomagnetic storms, scattering of galactic cosmic rays and produce other perturbations of all near-Earth environment. Thus, detection of the CME initiation in the solar corona and prediction of its propagation in the heliosphere is one of the most important tasks of the space weather forecasting. Commonly such prediction consists in estimation of arrival time and speed of the CME frontal structure to the observation site without taking into account the post-eruption effects. However, very often \citep[see, e.g.][]{lugaz2017,rodkin2018}, due to their large-scale structure, successive CMEs may interact in the heliosphere, which results in a change of their initial kinematic and magnetic parameters. A statistical study of \citet{temmer2017} in the time period 2011\,--\,2015 has shown that CMEs often are followed by a trailing region behind its rear with declining speed with duration up to several days which  much longer than the average CME duration itself (about 1.3 days). As a result, powerful CMEs can cause disturbances of the interplanetary medium, which may led to significant deviations of the CME arrival times and speeds from the initially predicted, especially, in the case of slow CMEs \citep{mostl2014,coronaromero2017,shugay2018,ravish2019}.

The aim of this work is to clarify the nature of the post-Interplanetary Coronal Mass Ejection (ICME) perturbation of SW: whether it is associated with entrainment of the interplanetary medium by the passed CME or with some post-CME plasma flow from the eruption site. We consider the case of the large eruption occurred on 2010 August~18 at the western solar limb, which was observed by several space telescopes and coronagraphs, studied formation of the eruption flows in the corona and in the heliosphere and established the correspondence between the flows and the SW disturbances measured in situ. The state of the coronal plasma during the eruption process was determined from the differential emission measure (DEM) analysis performed on the base of the multiwave EUV images.The DEM function describes the amount of thermal plasma along the line of sight at a given electron temperature retrieved from intensities in spectral bands with different temperature responces. By separating emission in specific temperature ranges, DEM enables to discern the spatial and temporal dynamics of coronal structures participating in the eruption process \citep[e.g.,][]{grechnev2019, saqri2020, hein2021}. A number of algorithms have been developed to derive a coronal DEM from SDO/AIA images in multiple bandpasses \citep[e.g.,][]{hannah2012, plowman2013, plowman2020}. Detailed comparison of different algorithms is beyond the scope of this paper, and can be found, for example, in \citet{aschwan2015}. In our case of the limb eruption we use the method and software described recently in \citet{plowman2020}. Using the data determined by the DEM analysis (the plasma densities and emission-weighted temperatures), we performed modeling of the Fe-ion charge distribution of the plasma outflows ``frozen-in’’ at the boundary of the corona for several temporal intervals: in the quiet state before eruption, during the CME formation and after it’s liftoff. A comparison of the modeled Fe-ion charge states with the measured ones has shown that the post-ICME disturbance of SW is associated with the post-eruption coronal flow.

\section{Data} \label{sec:s1}

In the analysis of the eruption plasma in the inner corona at the distances up to 1.7~$R_\odot$ we used the EUV multiwavelength images from the Atmospheric Imaging Assembly telescope \citep[AIA:][]{lemen2012} aboard the Solar Dynamic Observatory (SDO), the images in the 174~{\AA} band of the Sun Watcher with Active Pixels and Image Processing (SWAP) telescope as a part of the Project for Onboard Autonomy~2 (PROBA2) mission \citep{seaton2013} and the images in 195~{\AA} band from the Extreme-Ultra-Violet Imager (EUVI) as a part of Sun Earth Connection Coronal and Heliospheric Investigation \citep[SECCHI:][]{howard2008} aboard the Solar TErrestrial RElations Observatory (STEREO) spacecraft \citep{kaiser2008}. At the distances 2\,--\,30~$R_\odot$ we explore the CME formation from the data of the LASCO C2 and C3 coronagraphs \citep{brueckner1995} aboard the Solar and Heliospheric Observatory \citep[SOHO:][]{domingo1995}. The SW data, including the Fe-ion charge distributions, magnetic field and plasma parameters, were taken from observations with the PLAsma and SupraThermal Ion Composition \citep[PLASTIC:][]{galvin2008} and the In situ Measurements of Particles And CME Transients \citep[IMPACT:][]{luhmann2008} instruments.

\section{Description of the event and kinematics of eruption} \label{sec:s2}

We investigated the eruption in active region (AR) 11093 on 2010 August~18, when it was seen from the Earth at the Western limb. During the preceding days, when this AR crossed the solar disk, it produced a series of eruptions studied by several researchers \citep{vemar2012,tun2013,lario2017,elke2017}. On August~18, the most powerful CME (partial halo) was registered above the Western limb by LASCO C2 at 05:48~UT and by STEREO-A/COR2 at 05:54~UT. The SolarDemon dimming detection system \citep{kra2015} determined that the eruption was accompanied by a large dimming that appeared in the corona above the limb at 05:02~UT and existed at least until 07:02~UT.

This eruption was observed in quadrature from two positions: from the Earth by SDO/AIA and Proba~2/SWAP EUV telescopes, by SOHO/LASCO coronagraphs, and from the STEREO-A position by the SECCHI/EUVI telescope. The ICME and subsequent post-eruption SW flows were detected in situ by the STEREO-A/PLASTIC and IMPACT instruments.

\begin{figure}[htb!]
\plotone{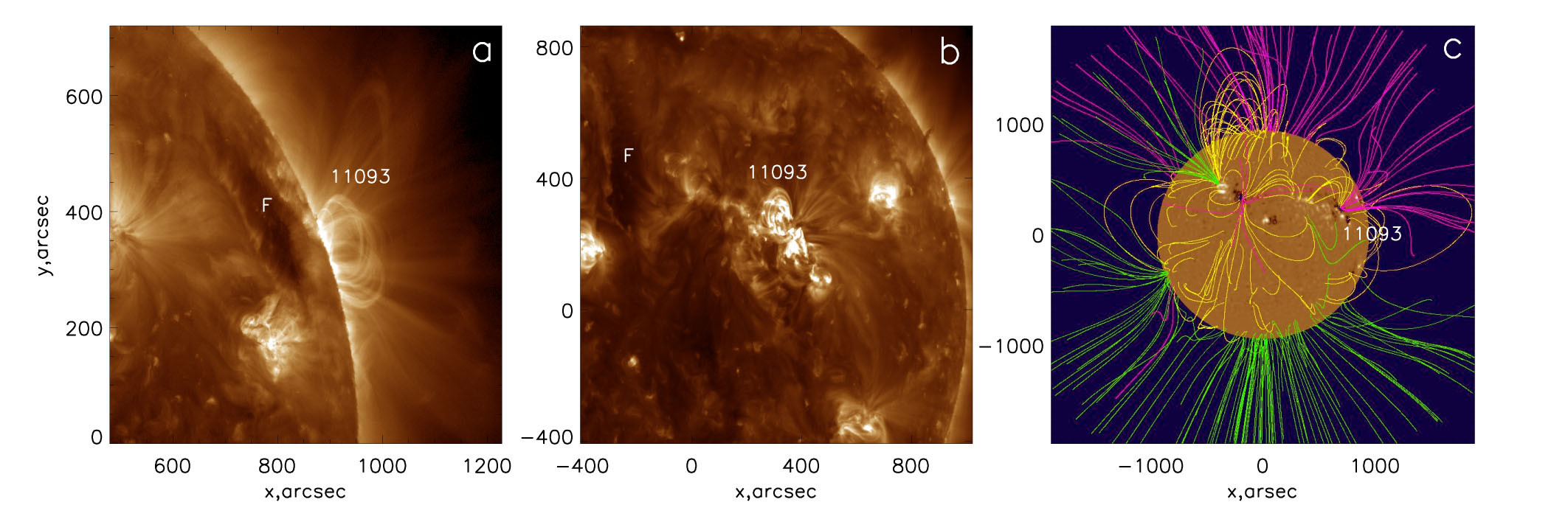}
\plotone{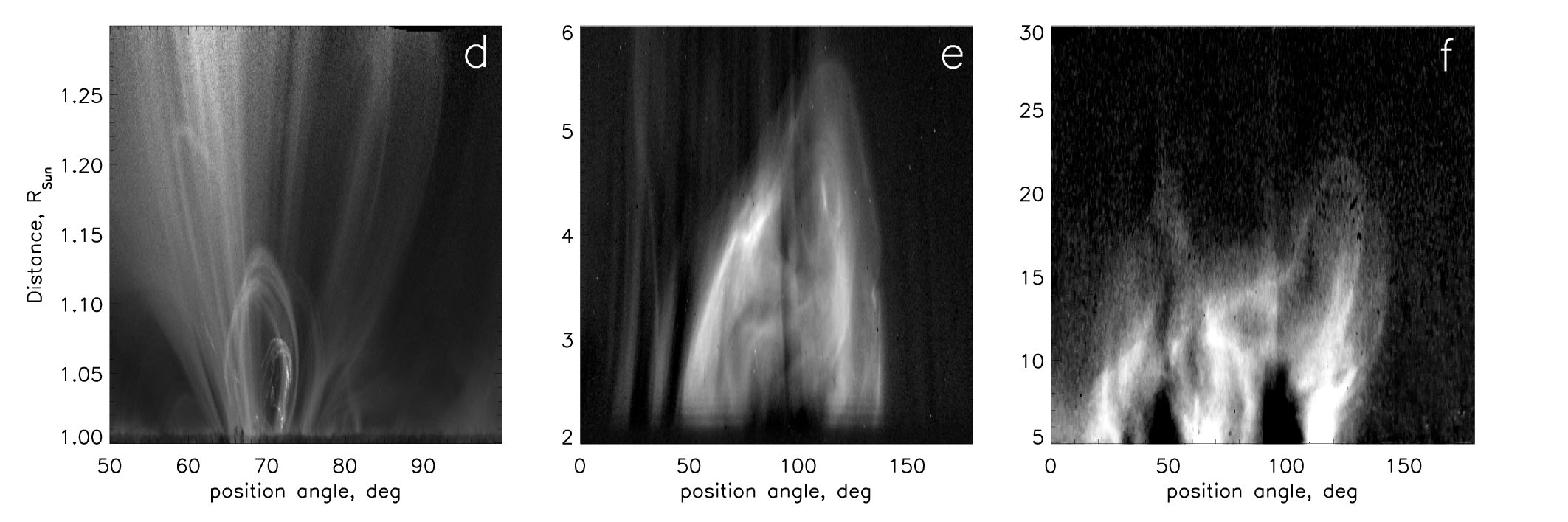}
\plotone{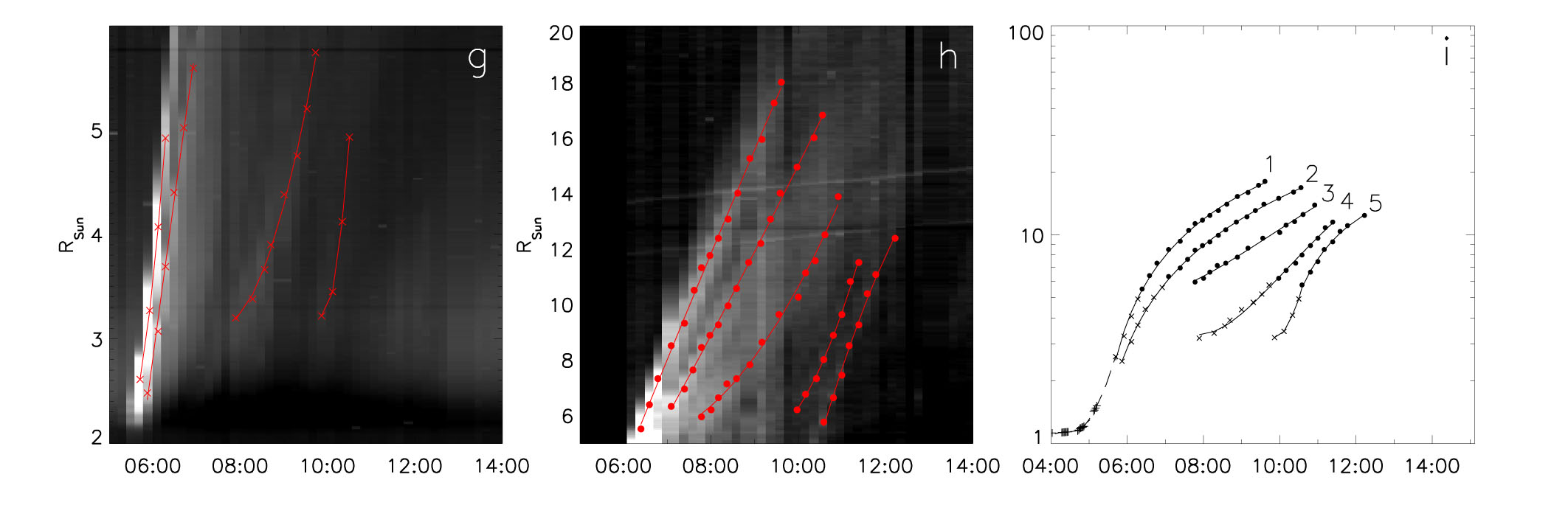}
\caption{(a) Active Region 11093 seen on 2010 August 18 before eruption by SDO/AIA in 193~{\AA} at 04:00:07~UT; (b) by STEREO-A/EUVI in 195~{\AA} at 04:00:30~UT. (c) The magnetic field map of the Sun in the PFSS approximation for 2010 August 14, 07:00~UT (taken from the database \url{https://www.lmsal.com/solarsoft/archive/sswdb-new/packages/pfss/l1q_synop}). (d, e, f) The polar maps of the coronal structures during eruption in AIA 171~{\AA} (2010 August 18, 05:00:48~UT), LASCO C2 (06:12:07~UT) and LASCO C3 (09:06:05~UT). The position angle counts out clockwise from the North. (g) The time-distance (T-D) slice map of the eruption flows derived from the LASCO C2 polar maps; (h) the same from the LASCO C3 maps; (i) the combined T-D plots of the eruption flows including the AIA 171~{\AA} and SWAP 174~{\AA} (plus signs), C2 (crosses) and C3 (circles) data.
\label{fig:fig1}}
\end{figure}

In Figure~\ref{fig:fig1} the images of AR~11093 are shown before eruption on August~18 at 04:00~UT in the AIA 193~{\AA} band (Figure~\ref{fig:fig1}a) and in the STA/EUVI 195~{\AA} band (Figure~\ref{fig:fig1}b). Figure~\ref{fig:fig1}c shows the PFSS-modeled magnetic field structure on 2010 August~14 at 07:00~UT, when the AR was on the solar disk, taken from the Lockheed Martin Solar and Astrophysics Laboratory (LMSAL) archive\footnote{\url{https://www.lmsal.com/solarsoft/archive/sswdb-new/packages/pfss/l1q_synop}}. The magnetic field structure included closed coronal loops and a pseudostreamer \citep{wang2015}, that separated a small equatorial coronal hole near AR~11093 and the northern polar coronal hole. Westward from the AR, a large filament F was seen, which later participated in the eruption.

At the initial stage of eruption, below a distance of 2~$R_\odot$, when the CME structure was not finally formed, the coronal loops started to move upward (Figure~\ref{fig:fig1}d) ,which was seen by the EUV SDO/AIA (up to 1.3~$R_\odot$) and PROBA2/SWAP (up to 1.7~$R_\odot$) telescopes. Then, from 2 to 5~$R_\odot$ (in LASCO~C2 observations), during the solar flare (GOES C4.5, start at 04:45~UT, maximum at 05:48~UT, end at 09:30~UT), the eruption plasma was accelerated and transformed from a system of loops to several successive compact structures. At that period, the filament F seen in Figure~\ref{fig:fig1}a and \ref{fig:fig1}b manifested a draining towards the eruption site along the neutral magnetic line, so it might enrich the eruption plasma by the cold filament material.

Figure~\ref{fig:fig1}d, \ref{fig:fig1}e and \ref{fig:fig1}f demonstrate the polar intensity maps of AIA 171~{\AA} (2010 August 18, 05:00:48~UT), LASCO C2 (06:12:07~UT) and LASCO C3 (09:06:05~UT). The position angle counts out clockwise from the North (it corresponds to the ordinary LASCO position angle by the relation $360^\circ$ -- PA$_{LASCO}$). The apex of the coronal structure seen at the latitude angle of $73^\circ$ in EUV below 1.3~$R_\odot$ shifted to $115^\circ$ at $R=2-5~R_\odot$ (C2) and split into three parts above 10~$R_\odot$ (C3).

We identified the upward plasma flows by two methods. The confined plasma flows were identified as bright ridges on the time-height maps (Figure~\ref{fig:fig1}g and \ref{fig:fig1}h) created from the LASCO C2 and C3 (Figure~\ref{fig:fig1}e and \ref{fig:fig1}f) polar maps.

To select the plasma flows that may contribute to SW, we integrated intensity of the polar maps at each height on the C2 map within the position angles $76.3^\circ - 86.3^\circ$  and on the C3 map within $86.3^\circ\pm2^\circ$. These ranges correspond to inclination of the eruption structure with distance to the STA position near the equator (the HEEQ latitude of STA for 2010 August~18 was equal to $3.7^\circ$). According to the WSA-ENLIL-DONKI-HELCATS model \footnote{\url{http://helioweather.net/archive/2010/08/}}, in the ecliptic plane the CME angular width is more than $60^\circ$, whereas the STA position is declined from the CME apex on the angle less than $10^\circ$. Thus, no corrections for the CME geometry and projection effect have been needed. As a result, we obtained the time-distance maps shown on Figure~\ref{fig:fig1}g (the C2 map) and Figure~\ref{fig:fig1}h (the C3 map). We distinguished 4 flows on the C2 map and 5 flows on the C3 map which are combined together on Figure~\ref{fig:fig1}i. The flows 1 and 2 seen in C2 and C3 are evidently match with the initial upward moving structure seen in EUV below 1.3~$R_\odot$. The flows 3, 4 and 5 are seen only on the C2 and C3 maps, so they were originated above 2~$R_\odot$, probably at the streamer’s top. The visibility of all flows diminishes with distance so, that they become indistinguishable from background between 10 and 20~$R_\odot$ due to weakening of contrast. Nevertheless, we expected that all flows may appear in SW.

All five flows were originated and accelerated in the corona during the C-class flare up to 09:30~UT. After the end of the flare, the LASCO images showed that the plasma outflow lasted as a continuous structureless stream. We estimated the speed of this continuous flow by analysis of cross-correlation between irregularities of intensity in the C3 polar maps at different heights as a function of time between 11:18~UT and 12:54~UT. We found the mean speed of this component to be $\backsim 530~km~s^{-1}$, which corresponds to its startup from the solar surface at about 10:00~UT. We regard this component as the sixth flow.

We suggested that after the end of the flare the flows were decelerated by two forces: the gravitation force below 20~$R_\odot$ and the MHD drag force above that distance \citep{shen2012, grechnev2019}. The gravitational deceleration is described by the formula:
\begin{equation}
v_1^2 = v_0^2 - 2GM_\odot(\frac{1}{r_0} - \frac{1}{r_1})\, ,
\end{equation}
with $v_0 (v_1)$ being a velocity of the flow at the distance of $r_0 (r_1)$ from the Sun’s center, G is the universal gravity constant, and $M_\odot$ is the mass of the Sun. In the Drag-based Model \citep[DBM:][]{vrsnak2013, vrsnak2021} deceleration is defined by the magnetic drag force between the flow and the ambient media according to the equation:
\begin{equation}
dv/dt=-\gamma(v-w)|v-w|\, ,
\end{equation}
where $v$ is the flow speed, $w$ is the speed of the ambient SW, $\gamma$ is the MHD drag parameter. Integration of this equation along the path from the initial point to the observer gives the arrival speed and time. As the initial parameters, the model uses the flow speeds and times at $R=20~R_\odot$.

The equation~(2) is basically analogous to the aerodynamic drag considered by \citet{carg2004}. The drag parameter can be expressed as:
\begin{equation}
\gamma=\frac{c_d}{L\cdot(\frac{\rho}{\rho_w}+\frac{1}{2})}\, ,
\end{equation}
where $c_d$ is the dimensionless drag coefficient (typically, $c_d=1$), $L$ is the thickness of the ICME in the radial direction, $\rho$ and $\rho_w$ are densities of the ICME and the ambient plasma. In the typical conditions, $\gamma=0.2 – 2.10^7~km^{-1}$ (for details, see \citet{vrsnak2013, zic2015}).

As it follows from the definitions, the model was developed for a single isolated CME propagated in the stationary ambient solar wind. The advanced version of DBM -- the Drag-Based Ensemble model \citep{dumb2018} uses the most probable model parameters based on the typical CME geometry and ambient plasma conditions, which is important for operative forecasting. Examples of application of DBM and comparison with other prediction models can be found in \citet{vrsnak2014, shi2015, napol2018, vrsnak2021, dumb2021} and references herein. The drag-based model for prediction of the CME fronts and sheaths directed to the Earth  using the height dependence and de-projected velocities with assistance of the STEREO data was described by \citet{hess2015}. The DBM-based models for prediction of the Earth-directed CMEs using the data from the STEREO Heliospheric Imagers are recently developed, such as the elliptic model ElEvoHI \citep{rollett2016} and ElEvoHI 2.0 elliptic front deformation model \citep{hint2021}.

In our case, the CME is formed from a series of consecutive flows followed by the unstructured post-eruption flow propagated towards STA, so we cannot use the STEREO data about their geometric shape and movement outside the corona. However, we suggested that DBM can be used in this case taking into account that the flows propagated along the same magnetic channel opened by the frontal structure of the CME with the appropriate ambient plasma speed. By the use of DBM, we aimed to check its applicability for this non-typical case and determine the particular values of the model parameters for the best agreement of the model results with the measurements. As the decisive parameter, we consider matching of the modeled flow speed with the measured in situ speed of the SW protons for the whole ensemble of flows.

Table~\ref{tab:tab1} presents the initial data and the modeled times and speeds of the SW flows calculated with the base version of DBM in comparison with the measured in situ on STA. We have made a series of calculations and found that the discrete flows 1 -- 3 and 5 well correspond to the ICME density peak with $\gamma=0.3\cdot10^{-7}~km^{-1}$ for the first flow and $0.13\cdot10^{-7}~km^{-1}$ for the others (the value averaged over the flows 2 -- 5), and the ambient plasma speed for all flows $w=445~km\cdot s^{-1}$. For the unstructured flow we found $\gamma=0.13\cdot10^{-7}~km^{-1}$ and the same speed of the ambient plasma.

The spreads in the flow arrival times and speeds are summarized from two parts: uncertainties from the time-distance maps due the data discretization shown in the $T_{20}$ and $V_{20}$ columns of Table~\ref{tab:tab1}, and the model uncertainties of DBM. Typically, DBM gives the CME arrival times with the uncertainty of 9 -- 14 hours (see the references cited above). However, our investigation based on the data of 2010 -- 2011 \citep{rodkin2018} has shown that in the case of the single ICMEs (not interacted with other SW transients in the heliosphere) the inaccuracy of the DBM results in the arrival time amounts of 8 hours, in the speed -- $65~km\cdot s^{-1}$. As a result, for all flows, except flow 4, the difference between the modeled and measured times and speeds did not exceed the final errors in $V_{STA}$ in Table~\ref{tab:tab1}. Flow 4 did not fit to the SW data under any calculation parameters, probably, because it was rather weak and merged to the main flows.

\begin{deluxetable*}{cccccccc}[htb!]
\tablenum{1}
\tablecaption{Kinematic parameters of the CME and post-eruption flows \label{tab:tab1}}
\tablewidth{0pt}
\tablehead{
\colhead{№ of} & \colhead{$R_{td}$} & \colhead{$T_{td}$} & \colhead{$T_{20}$} & \colhead{$V_{20}$} & \colhead{$T_{STA}$}  &  \colhead{$V_{STA}$} & \colhead{$V_{p}$} \\
\colhead{flow} & \colhead{$R_\odot$} & \colhead{Aug 18 (UT)} & \colhead{Aug 18 (UT)} & \colhead{($km~s^{-1}$)} & \colhead{(UT)} & \colhead{($km~s^{-1}$)} & \colhead{($km~s^{-1}$)}}
\startdata
1 & 17.75 & 09:36 & $10:15\pm0.22h$ & $670\pm26$ & Aug 21 00:58$\pm10h$ & $533\pm66$ & 585 \\
2 & 16.74 & 10:33 & $11:37\pm0.24h$ & $599\pm23$ & Aug 21 03:20$\pm12h$ & $550\pm69$ & 580 \\
3 & 13.59 & 10:55 & $12:42\pm0.28h$ & $686\pm32$ & Aug 20 22:27$\pm12h$ & $590\pm70$ & 551 \\
4 & 12.30 & 11:23 & $13:17\pm0.33h$ & $872\pm69$ & Aug 20 12:59$\pm12h$ & $671\pm75$ & 329 \\
5 & 12.47 & 12:13 & $14:36\pm0.37h$ & $606\pm48$ & Aug 21 05:20$\pm18h$ & $553\pm80$ & 574 \\
6 & 9.38 & 12:54 & $16:53\pm0.40h$ & $508\pm56$ & Aug 21 19:39$\pm16h$ & $500\pm77$ & 511 \\
\enddata
\tablecomments{$R_{td}$ is a maximum distance in the time-distance map, $T_{td}$ is a time at $R_{td}$, $T_{20}$ is a time at $20R_\odot$, $V_{20}$ is a speed at $20R_\odot$, $T_{STA}$ is an arrival time at STA modeled by DBM, $V_{STA}$ is a speed at STA modeled by DBM, $V_p$ is a speed of protons measured at STA}
\end{deluxetable*}

\section{DEM diagnostics of plasma flows in the corona} \label{sec:s3}

We investigated variation of the plasma temperature and density in the eruption region using the DEM distribution derived from the AIA EUV images. Intensity fluxes in spectral bands are related to DEM via the expression:
\begin{equation}
F_i=\int_T G_i(T)\mathrm{DEM}(T) \,dT \, ,
\end{equation}
where $F_i$ is the intensity flux and $G_i(T)$ is the temperature response function of the passband i.

To determine DEM for the eruption plasma, we applied the method and software developed by \citet{plowman2020}. The integral of DEM over the temperature gives the total emission measure (EM) integrated along the line of sight: $\mathrm{EM}=\int_T \mathrm{DEM}(T) \,dT$.
Using the obtained DEM, one can calculate the emission-weighted temperature \citep[e.g.,][]{cheng2012,saqri2020}:
\begin{equation}
T_{\mathrm{em}} = \frac{\int_T T\cdot \mathrm{DEM}(T) \,dT}{\mathrm{EM}} \, .
\end{equation}

Based on the EM structure, we can estimate the plasma density, assuming that the depth of the structure along the line of sight ($L$) is approximately equal to its visible width \citep[the same method can be found, for example, in][]{cheng2012}. So the plasma density can be estimated as :
\begin{equation}
N_e=\sqrt{\frac{\mathrm{EM}}{L}}\, .
\end{equation}

To retrieve DEM, we used the AIA/SDO images of the solar corona in six channels (94~{\AA}, 131~{\AA}, 171~{\AA}, 193~{\AA}, 211~{\AA}, 335~{\AA}), that coverage a broad temperature range (from $10^5$ to above $10^7K$). The input error of each pixel in AIA images was calculated using $aia_-bp_-estimate_-error$ routine considering the data obtained during eclipses \citep{hein2021}.

Stray light in some cases can significantly disturb intensities of the AIA images \citep{wend2018,saqri2020,hein2021} in particular, above the limb, where intensities are fast weakening. In our case no correction for stray light was needed within the typical error of the DEM reconstruction (about 20~\%).

The errors may occur in the DEM solution at high flow speeds (over $800~km~s^{-1}$, see \citet{grechnev2019}) due to non-simultaneity in the registration of images in different AIA channels. In our case, it is possible to compute the DEM without a compensation for its motion, since the flow speeds at distances $1.2~R_\odot$ were about $100~km~s^{-1}$.

We analyzed DEM in the AIA field of view for the several moments on 2010 August 18: before the eruption at 04:01~UT, during the development of the solar flare at 05:20~UT, 05:22~UT and 05:25~UT, when the plasma rose up in the AIA field of view, and in the post-eruption stage at 10:00~UT, when the CME leaved the corona.

\begin{figure}[htb!]
\plotone{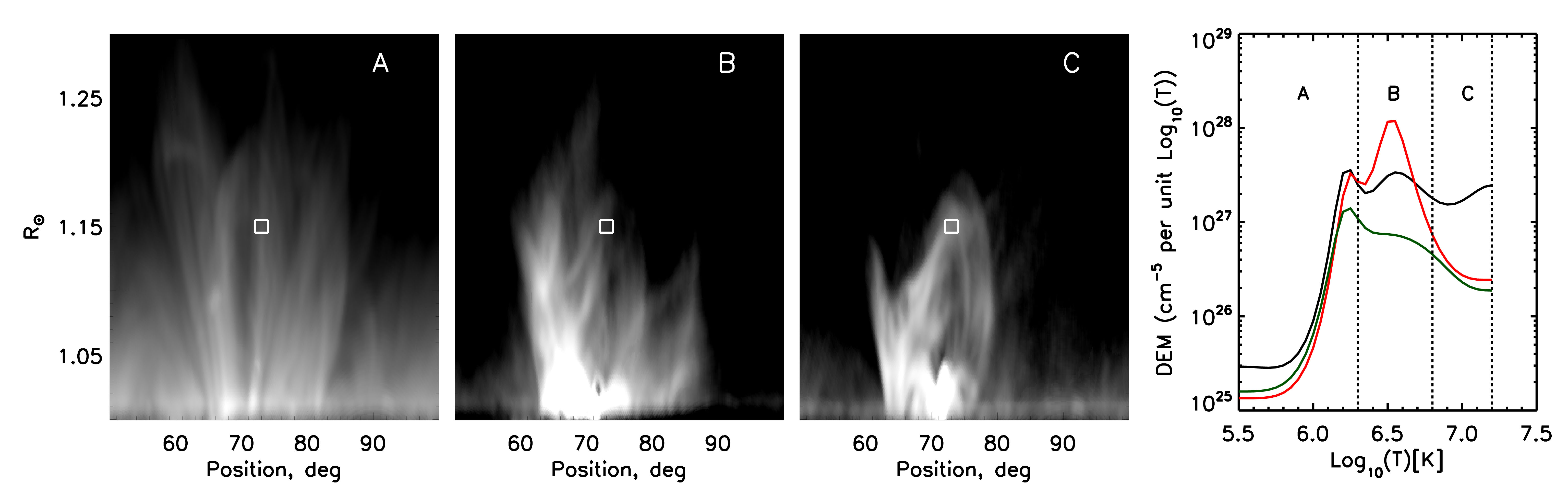}
\caption{Polar EM maps for 3 temperature ranges (A: $5.5 < \mathrm{log}_{10}T < 6.3$, B: $6.3 < \mathrm{log}_{10}T < 6.8$, and C: $6.8 < \mathrm{log}_{10}T < 7.2$) at 05:20~UT and the DEM temperature distribution in the indicated box at three time moments. Red curve -- 04:01 UT ($\chi^2$ = 0.99), black -- 05:20~UT ($\chi^2$ = 0.95), green -- 10:00~UT ($\chi^2$ = 0.94)  \label{fig:fig2}}
\end{figure}

In Figure~\ref{fig:fig2}, we show the EM polar maps at 05:20~UT in three temperature ranges $5.5 < \mathrm{log}_{10}T < 6.3$ (A), $6.3 < \mathrm{log}_{10}T < 6.8$ (B), and $6.8 < \mathrm{log}_{10}T < 7.2$ (C). The right panel shows variation of the DEM temperature distribution in the indicated box before (04:01~UT), during (05:20~UT) and after (10:00~UT) eruption. The DEM profiles were averaged over the box $1.15\pm0.02 R_\odot$ and $73^{\circ}\pm1^{\circ}$ in the position angle, which corresponds to the highest total EM during eruption. The DEM was considered only up to $\mathrm{log}_{10}T = 7.2$ because of the artifacts appeared due to the low temperature sensitivity of the AIA channels at high temperatures \citep{plowman2020}. To calculate the plasma density in the hottest range ($\mathrm{log}_{10}T>6.8$), the value of EM integrated over the temperature was doubled, since the DEM distribution at this temperature range included only half of the peak.

In Table~\ref{tab:tab2} we show the mean values of plasma density and emission-weighted temperature in three ranges: $5.5 < \mathrm{log}_{10}T < 6.3$, $6.3 < \mathrm{log}_{10}T < 6.8$, and $6.8 < \mathrm{log}_{10}T < 7.2$. The mean values at 05:20~UT, 04:01~UT, 10:00~UT were averaged over the box as described above. The box was shifted to $1.17~R_\odot$ at 05:22~UT, and to $1.20~R_\odot$ at 05:25~UT according to the plasma movement. In the quiet conditions, before eruption (04:01~UT) and after eruption (10:00~UT), densities in the hot temperature range ($6.8 < \mathrm{log}_{10}T < 7.2$) were larger than those in the cold range ($5.5 < \mathrm{log}_{10}T < 6.3$), although in the histograms of Figure~\ref{fig:fig2} the relation between the corresponding EM values is reverse. The cause is that in the quiet conditions the depth of the hot structures was much less (more than one order) than of the cold ones. During the eruption (05:20\,--\,05:25~UT), density in the hot range increases due to inflow of the heated plasma of the reconnected surrounding loops with the medium temperature $6.3 < \mathrm{log}_{10}T < 6.8$. After eruption (at 10:00~UT) densities in all temperature ranges drop down due to liftoff of the CME.

\begin{deluxetable*}{ccccccc}[htb!]
\tablenum{2}
\tablecaption{The mean values of plasma density and emission-weighted temperature for 5 time moments in three temperature ranges \label{tab:tab2}}
\tablewidth{0pt}
\tablehead{
\colhead{} & \multicolumn{2}{c}{$5.5 < \mathrm{log}_{10}T < 6.3$} & \multicolumn{2}{c}{$6.3 < \mathrm{log}_{10}T < 6.8$ } & \multicolumn{2}{c}{$6.8 < \mathrm{log}_{10}T < 7.2$ } \\
\cline{2-7}
\colhead{Time} & \colhead{$N_e~(10^8)$} & \colhead{$T_{\mathrm{em}}$} & \colhead{$N_e~(10^8)$} & \colhead{$T_{\mathrm{em}}$} & \colhead{$N_e~(10^8)$} & \colhead{$T_{\mathrm{em}}$} \\
\colhead{(UT)} & \colhead{$(cm^{-3})$} & \colhead{$(MK)$} & \colhead{$(cm^{-3})$} & \colhead{$(MK)$} & \colhead{$(cm^{-3})$} & \colhead{$(MK)$}
}
\startdata
04:01 & $1.0\pm0.1$ & $1.5\pm0.4$ & $4.8\pm0.5$ & $3.4\pm0.7$ & $3.7\pm0.2$ & - \\
05:20 & $1.3\pm0.1$ & $1.4\pm0.3$ & $3.3\pm0.2$ & $3.6\pm0.3$ & $5.2\pm1.0$ & $10.6\pm4.5$ \\
05:22 & $1.1\pm0.1$ & $1.4\pm0.2$ & $3.1\pm0.2$ & $3.6\pm0.4$ & $4.6\pm1.0$ & $10.5\pm4.9$ \\
05:25 & $1.0\pm0.1$ & $1.4\pm0.3$ & $2.5\pm0.2$ & $3.5\pm0.5$ & $3.8\pm0.5$ & $10.6\pm3.2$ \\
10:00 & $0.8\pm0.1$ & $1.4\pm0.1$ & $2.0\pm0.4$ & $3.3\pm0.9$ & $2.5\pm0.6$ & - \\
\enddata
\end{deluxetable*}

\section{Solar wind parameters measured in situ} \label{sec:s4}

\begin{figure}[htb!]
\centerline{
\includegraphics[width=0.7\textwidth,clip=]{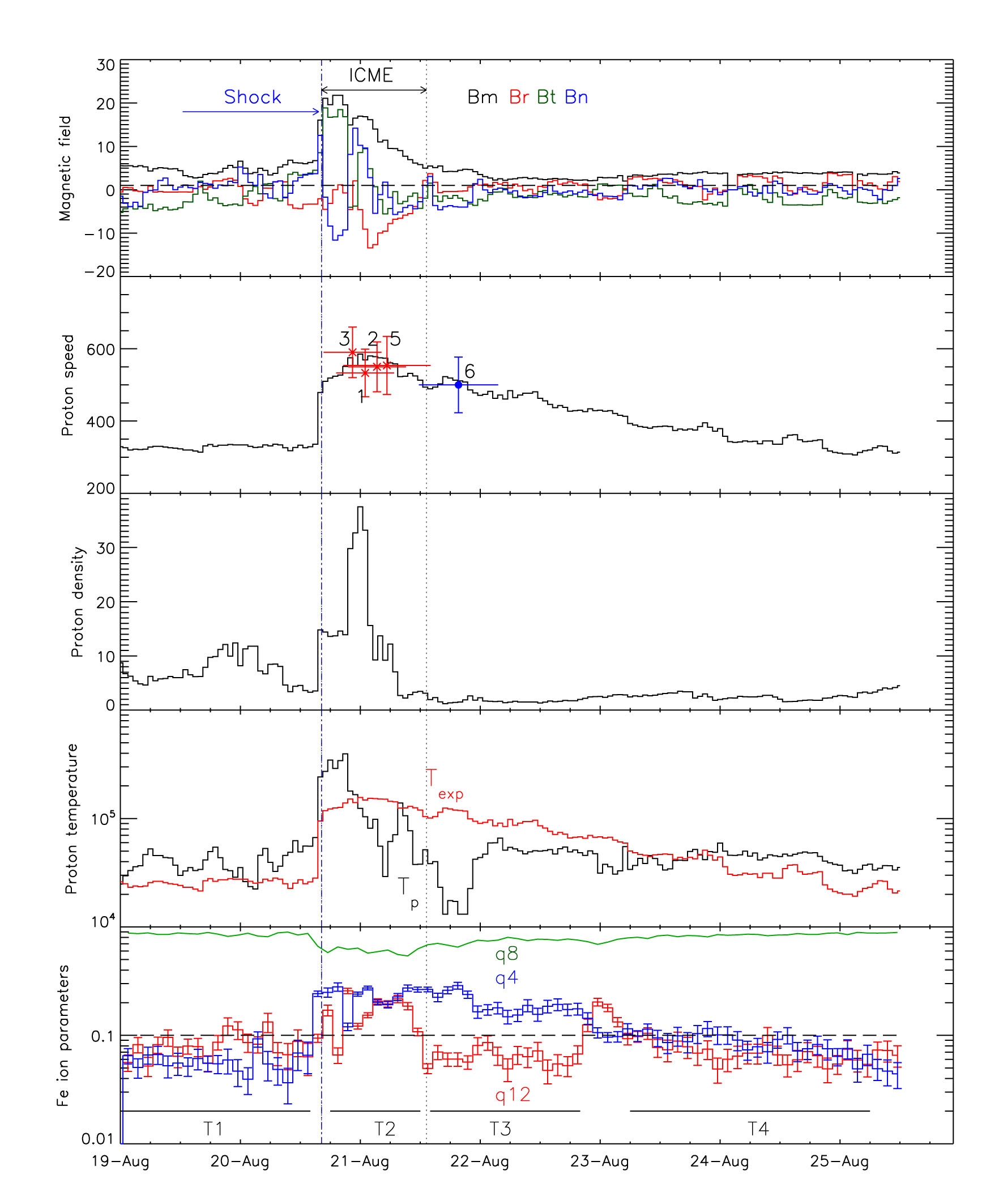}
}
\caption{Parameters of the magnetic field, proton speed, density, kinetic temperature and Fe-ion distributions of SW from measurements by PLASTIC and IMPACT at STA in comparison with the results of calculations of the flow arrival times and speeds by the Drag-based model. The dotted lines designate the start and end of the ICME according to the STEREO-A ICME catalog (\url{https://stereo-ssc.nascom.nasa.gov/data/ins_data/impact/level3/ICMEs.pdf}).
\label{fig:fig3}}
\end{figure}

Figure~\ref{fig:fig3} shows the SW parameters registered by IMPACT and PLASTIC at STA in the period 2010 August~19\,--\,26: the magnetic field magnitude and its RTN components, proton speed, density and temperature, and parameters q4, q8 and q12 (see the definition of these parameters below in equation~(7)) characterizing the Fe-ion charge distribution. The SW disturbance started with the shock on 2010 August~20, 16:14~UT\footnote{\url{http://ipshocks.fi/database}}, the ICME on August~20, 16:14~UT -- August~21, 13:14~UT (from the STEREO event catalog \footnote{\url{https://stereo-ssc.nascom.nasa.gov/data/ins_data/impact/level3/ICMEs.pdf}}) followed up to August~25 by a long trailing region with the declining speed. The arrivals of the plasma flows given in Table~\ref{tab:tab1} are marked on the proton speed chart. The flows 1, 2, 3 and 5 evidently correspond to the magnetic cloud of the ICME. The forth flow (not shown on the chart) with the highest speed according to Table~\ref{tab:tab1} should arrive significantly ahead of the ICME, but it was not observed in SW. Most likely, this weak flow decelerated and merged with other discrete flows. The sixth post-eruptive flow arrived in the post-ICME part of the transient. It is worth to mention, that the shock time well agrees with the DBM modeling of the CME frontal structure indicated in the LASCO Coordinated Database Analysis Workshops (CDAW) database (start at 05:48~UT, $V_{20}=1416~km~s^{-1}$) with $\gamma=0.3\cdot 10^{-7}$ and $w=330~km~s^{-1}$.

The ionization state of the SW plasma is ``frozen-in’’ in the inner corona at heights where recombination/ionization timescales become dominant over the plasma expansion timescale \citep[see, e.g.,][]{hund1968, ko1997, goryaev2020}. Since the Fe ions freeze-in at the largest heights than other abundant elements, the Fe-ion composition is the most suitable for characterizing the state of the SW plasma. In the development of the approach proposed in \citet{goryaev2020}, we introduce here three relative parameters q4, q8, and q12 for characterizing the Fe-ion charge distribution as follows:
\begin{equation}
q4 = \frac{\sum_{0\leq Z\leq 7} n_z}{\sum_{0\leq Z\leq 20} n_z},\: q8 = \frac{\sum_{8\leq Z\leq 11} n_z}{\sum_{0\leq Z\leq 20} n_z},\: q12 = \frac{\sum_{12\leq Z\leq 20} n_z}{\sum_{0\leq Z\leq 20} n_z},\; q4+q8+q12 = 1 \, ,
\end{equation}
where $n_z$ is the number density of the ion with charge $Z$. The parameters q4, q8, and q12 correspond conditionally to the ``cold'', ``middle'', and ``hot'' parts of the Fe charge distribution.

In contrast to the average charge $Q_{\mathrm{Fe}}$, the differentiation for the parameters q4, q8, and q12 shown in Figure~\ref{fig:fig3} allows one to derive a more detailed information on conditions in the SW plasma.

In the period from August~19 to the shock time the Fe-ion charge distribution was characterized by the largest value of q8$>$0.8 and the minor q4 and q12 values below 0.1, which corresponds to the slow SW.  After the shock, the jump of q4 to 0.3 indicated appearance of the sheath cold matter, then, in the ICME part, arrival of the hot eruption plasma has led to intermittent rise up of q12 to 0.2\,--\,0.3. The most interesting anomaly in the Fe-ion charge distribution is seen in the period after the rear of the ICME, from August~21, 13:14 UT to August~23, 00:00~UT. During this post-ICME period, the values of q4 were several times higher than that of q12, which means a significant domination of the ``cold’’ component in the SW plasma. Development of the dimming in the period from the shock to August 23, 12:00~UT was indicated by a decrease of the medium temperature parameter q8. After August 23, 12:00~UT all charge distribution parameters returned to the slow SW values.

\section{Fe-ion charge distributions of the flows in the corona and in SW around the ICME event} \label{sec:s5}

For our analysis of the ICME event, we considered the four time intervals designated as T1\,--\,T4 in Figure~\ref{fig:fig3}: 2010-08-19 00:00 -- 2010-08-20 14:00 (T1); 2010-08-20 18:00 -- 2010-08-21 12:00 (T2); 2010-08-21 14:00 -- 2010-08-22 20:00 (T3); and 2010-08-23 06:00 -- 2010-08-25 06:00 (T4). The interval T1 corresponds to the pre-ICME SW, T2 is associated with the passage of the ICME, T3 is related to the post-eruptive SW flows, and T4 is SW returned to pre-ICME conditions. For each time interval we summed the 2 hour Fe-ion distributions taken from the STEREO-A/PLASTIC database.

\begin{figure}[htb!]
\plotone{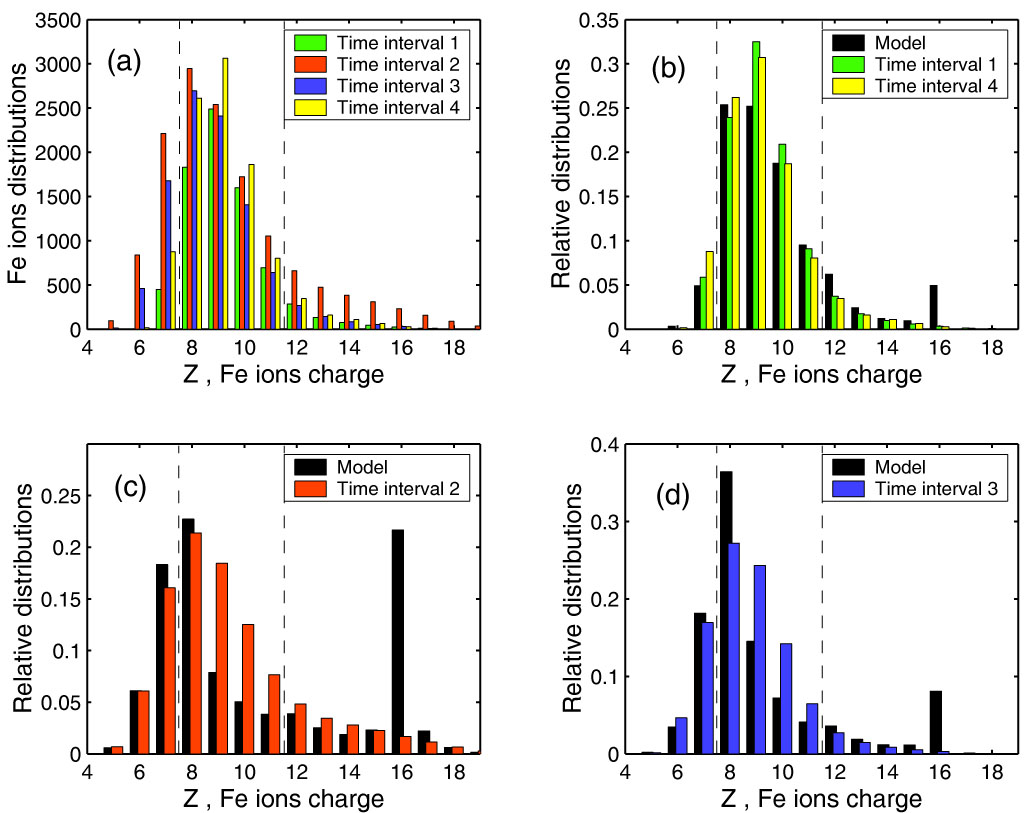}
\caption{The Fe-ions distributions for four time intervals associated with the ICME event as compared with modeled ones (in black). (a) The summed Fe-ion distributions (number of counts of charged Fe ions) for four time intervals T1\,--\,T4; (b) the relative (normalized to unity) distributions for time intervals 1 (green) and 4 (yellow); (c) the relative distribution for the time interval 2 (red); (d) the relative distribution for the time interval 3 (blue). The vertical dashed lines separate the ranges of ionic charge Z corresponding to the parameters q4 (left), q8 (middle), q12 (right).  \label{fig:fig4}}
\end{figure}

Figure~\ref{fig:fig4}a shows the number of counts of Fe ions with the charge $Z$ summed over the intervals T1\,--\,T4. The distributions for the intervals T1 (green) and T4 (yellow) have similar shapes peaked at $Z=9$, because these are probably associated with the slow component of SW. Though the distributions for T2 (red) and T3 (blue) intervals are associated with different SW types, these are peaked at the same value $Z=8$. Also, T2 distribution has the noticeable high charge tail with $Z\geq 12$, which is an evidence for the ICME event.

We modeled the evolution of the Fe-ion charge state and ``frozen-in’’ conditions for the CME event on 2010 August~18 to compare with the in situ observations of the associated ICME. To determine a ``frozen-in’’ charge state, we solved a system of balance equations with the preset profiles of plasma electron temperature $T_e$ and density $N_e$, and bulk velocity $V$ as functions of distance \citep[see details in][]{rodkin2017, grechnev2019}. We have performed three sets of calculations for the initial times from Table~\ref{tab:tab2}: 04:01~UT (hereafter case~1), 05:20~UT (case~2), and 10:00~UT (case~3). The bulk velocities $V$ for three cases were derived from the kinematic measurements. For the electron densities $N_e$ we used the plausible geometry of the expansion of the SW plasma in the corona: $N_e\sim 1/h^{2}$ ($h$ is the height above the solar surface) for the cases~1 and 3 (the slow SW flowing from streamers), and $N_e\sim 1/r^{3}$ ($r$ is the heliocentric distance) for the case~2 (CME flux rope). For the plasma temperatures we used the initial measured values from Table~\ref{tab:tab2} and performed a fit for the temperature profile to have a good agreement with the in situ observations. A similar procedure was applied by \citet{landi2012} to the coronal hole and equatorial streamer model. Our calculations have shown that our models for plasma densities and temperatures agree well with the results of \citet{landi2012} for cases~1 and 3 and with numerical MHD simulations of \citet{lynch2011} for the CME flux rope (case~2).

For deriving the final ``frozen-in’’ Fe-ion charge distributions in three mentioned cases we used the following procedure. According to Table~\ref{tab:tab2}, the SW flow for each case consists of two (cases~1 and 3) or three (case~2) plasma components, which hereafter are called ``cold’’, ``middle’’, and ``hot’’ ones according to temperature regimes. For each case, the ``frozen-in’’ charge distributions are then calculated separately for all components, and the final total distribution ($n_z$ values for all $Z$ under consideration) is given by mixing of plasma components:
\begin{equation}
n_z=\frac{n_z^{(c)}N_e^{(c)}L^{(c)} + n_z^{(m)}N_e^{(m)}L^{(m)} + n_z^{(h)}N_e^{(h)}L^{(h)} }{ N_e^{(c)}L^{(c)}+ N_e^{(m)}L^{(m)}+ N_e^{(h)}L^{(h)}} \, ,
\end{equation}
for case 2 (CME plasma) and
\begin{equation}
n_z = \frac{n_z^{(c)}N_e^{(c)}L^{(c)} + n_z^{(m)}N_e^{(m)}L^{(m)} }{ N_e^{(c)}L^{(c)}+ N_e^{(m)}L^{(m)}} \, ,
\end{equation}
for cases 1 and 3 (without the hot components), where $n_z^{(c)}$, $n_z^{(m)}$, $n_z^{(h)}$ are the partial final distributions for ``cold’’, ``middle’’, and ``hot’’ plasma components; $N_e^{(c)}$, $ N_e^{(m)}$, $ N_e^{(h)}$ and $L^{(c)}$, $ L^{(m)}$, $ L^{(h)}$ are the corresponding plasma densities and depths along the line of sight. We did not take into accont the hot component for the pre and post-eruption conditions due to large uncertaintes in the DEM hot wing at the temperatures $\mathrm{log}_{10}T>7$. A similar two-plasma model was used by \citet{grues2012} for possible interpreting ICME observations with very low and high charge state ions.

Figures~\ref{fig:fig4}b, \ref{fig:fig4}c, and \ref{fig:fig4}d show the comparison of the calculated relative (normalized to unity) Fe-ion distributions with the measured ones for the T1\,--\,T4 intervals. As it is seen, the modeled distributions agree well with the measurements. Nevertheless, the modeled distributions overestimate the high charge states with $Z\geq 16$ for all cases, especially for case~2 associated with the ICME. It may be related to the overestimation of the high temperature part of DEM distributions in the reconstruction procedure. The high charge peak for Fe$^{16+}$ ion is caused by the smaller recombination rates compared to the ions in lower charge states \citep[see][]{goryaev2020}. The comparison of q4, q8, q12 parameters for the modeled and measured Fe-ion distributions are shown in Table~\ref{tab:tab3}. It is also seen that q4 and q8 parameters have a very good agreement with the measured ones, while the modeled q12 values are overestimated about 2 times.

\begin{deluxetable*}{ccccccc}[htb!]
\tablenum{3}
\tablecaption{Measured parameters q4, q8, q12 for four time intervals 1--4 associated with the ICME event and post-ICME flow \label{tab:tab3}}
\tablewidth{0pt}
\tablehead{
\colhead{Time} & \multicolumn{2}{c}{q4} & \multicolumn{2}{c}{q8} & \multicolumn{2}{c}{q12} \\
\colhead{interval №} & \colhead{measured} & \colhead{modeled} & \colhead{measured} & \colhead{modeled} & \colhead{measured} & \colhead{modeled}}
\startdata
1 & $0.059\pm0.020$ & $0.053\pm0.013$ & $0.864\pm0.207$ & $0.789\pm0.094$ & $0.077\pm0.041$ & $0.158\pm0.107$ \\
2 & $0.229\pm0.035$ & $0.250\pm0.112$ & $0.600\pm0.083$ & $0.395\pm0.019$ & $0.171\pm0.040$ & $0.354\pm0.130$ \\
3 & $0.217\pm0.049$ & $0.218\pm0.016$ & $0.722\pm0.144$ & $0.623\pm0.095$ & $0.061\pm0.028$ & $0.159\pm0.111$ \\
4 & $0.089\pm0.028$ & - & $0.836\pm0.199$ & - & $0.074\pm0.038$ & - \\
\enddata
\tablecomments{The parameters for the time intervals 1--3 are compared with the modeled ones.}
\end{deluxetable*}

Furthermore, the modeling enables to interpret the difference in behaviour of the q4 and q12 parameters in the cases~1 (T1) and 3 (T3) (see Figure~\ref{fig:fig3}) as follows. In the pre-eruption state (case 1) the mean values of q4=0.059 and q12=0.077 corresponded to the ``frozen-in'' conditions, where the plasma transforms from collisional to collisionless state. It occurred at the heights of $h\approx4-5~R_\odot$ with the plasma electron temperature of $T_{e}\approx1~MK$ typical for the quiet slow SW. In the post-eruption state (case 3) q4 = 0.217, q12 = 0.061, and $T_{e}\approx0.5~MK$, so the ``cold’’ plasma dominates. According to the model results, this increase of the ``cold’’ component is explained not only by depletion of the highly charged ions after the CME liftoff, but also by the absence of heating after the flare ending. The last panel in Figure~\ref{fig:fig3} and data from Table~\ref{tab:tab3} shows that finally the SW plasma in the T4 interval returns to the pre-ICME state as in the T1 interval.

\section{Discussion and Conclusion} \label{sec:concl}

The analysis of the eruption in AR~11093 occurred on 2010 August~18 and the associated SW transient detected by STA on 2010 August~20 -- 23 has yielded the following results.

The CME flux rope was formed during the C-class flare from several discrete plasma flows originated in the region of AR~11093 with the pseudostreamer nearby. According to the AIA, SWAP EUV images and LASCO~C2\,--\,C3 time-height maps, the first flow started at $R=1.15~R_\odot$, others at $2.2~R_\odot$ and higher, probably, as the streamer blowout blobs like those observed recently by Parker Solar Probe \citep{lario2020, rouill2020}. After the end of the flare, the plasma outflow transformed into the unstructured stream, whose speed was determined by correlation of irregularities at different heights. To predict the arrival times and speeds of the flows at STA we applied the gravitation and DBM kinematic models. The drag parameters and ambient plasma speed were found in grid calculations under condition of minimal difference between the modeled flow speeds and the measured in situ proton speeds at the arrival times. Then the optimal values of the drag parameter and the ambient plasma speed were obtained by averaging the values for all flows except the fastest flow 4, which, probably, was decelerated and merged with other flows. As a result, we obtained that the discrete flows started during the eruption appeared in SW as components of the ICME, whereas the unstructured flow started after eruption appeared in the ICME trailing region. Thus, we showed that DBM is well applicable to the propagation of the multicomponent CME structure and also for the post-eruption flow. The ambient plasma speed was found to be of $445~km~s^{-1}$ for all flows (except flow 4), the $\gamma$ value was amounted to 0.3 for the frontal flow 1 and 0.13 for other flows. Such difference in $\gamma$ means that the drag force diminished for the flows passed behind the CME front along the same open magnetic field lines. The open magnetic field structure in the ICME trailing region was first described by \citet{neug1997}.

To understand the origins of the SW flows, we analyzed the Fe-ion charge distribution of SW and confronted it with parameters of the coronal plasma. First, we determined the plasma emission measure in the eruption region with the use of DEM in the time intervals corresponding to the pre-eruption, eruption and post-eruption conditions. The largest emission measure in all temperature bands was determined in the pre-eruption stage (at 04:01~UT). During the eruption (05:20\,--\,05:25~UT), the emission measure underwent on about one order in the middle temperature ($6.3 < \mathrm{log}_{10}T < 6.8$) and in the hot temperature ($6.8 < \mathrm{log}_{10}T < 7.2$) channels because of the uncompensated plasma outflow from the dimming region. In the cold channel ($5.5 < \mathrm{log}_{10}T < 6.3$) the EM value dropped down less than three times. Such lesser decrease can be explained by an additional inflow of the cold plasma by interchange reconnection of the cold loops seen in the AIA 171~{\AA} and SWAP 174~{\AA} with the open magnetic lines of the flux rope \citep{owens2020} and by drainage of the cold filament matter to the eruption site. Drainage of the filament mass before eruption was described by \citet{martin2008,bi2014,zhang2017,jenkins2018}.

Using the obtained plasma parameters, we have modeled the “frozen-in" Fe-ion charge distributions in the studied coronal flows and calculated the values of the q4, q8 and q12 parameters averaged over four time intervals T1\,--\,T4. For all the intervals, the modeled values agreed with the measured ones within the errors. A small excess of the measured q8 with respect to the modeled values may be related to contribution of the surrounding coronal structures not associated with the eruption. The observed anomalous divergence between the q4 and q12 values in the post-ICME trailing region can be explained by inflow of the cold plasma in absence of heating after the solar flare end. As a result, the SW ion state in the post-ICME tail in the interval 2010 August~21, 14:00~UT -- 2010 August~22, 20:00~UT significantly differs from the state of the slow ambient SW before 2010 August 20, 14:00 and after 2010 August~23, 06:00~UT.

Concluding, the results demonstrate that in the studied CME event of 2010 August~18\,--\,23 the post-ICME SW disturbance was caused by the post-eruption coronal flow from the dimming region and cannot be associated with the ambient SW entrained by the CME.

\begin{acknowledgments}
The authors thank the STEREO/PLASTIC, GOES, SDO/AIA, PROBA2/SWAP research teams for their open data policy. The PLASTIC investigation is an international (USA, Switzerland, Germany) effort involving the University of New Hampshire, the University of Bern, the Max-Planck-Institute for extraterrestrial Physics, Christian-Albrecht-University Kiel, NASA/Goddard Space Flight Center, and the University of California, Berkeley. The SDO image data are available by courtesy of NASA and AIA science team. SWAP is a project of the Central Spatial de Liege and the Royal Observatory of Belgium. This paper also uses data of the CME catalog that is generated and maintained at the CDAW Data Center by NASA and The Catholic University of America in cooperation with the Naval Research Laboratory. SOHO is a project of international cooperation between ESA and NASA. We are grateful for the opportunity to use the results of the simulation obtained by the Drag-Based Model.
\end{acknowledgments}

\bibliography{Formation_of_CME_and_post_eruptive_flow_ApJ}{}
\bibliographystyle{aasjournal}

\end{document}